%% file: main.tex
\documentclass[aps,prx,twocolumn,superscriptaddress,longbibliography]{revtex4-2}

\usepackage{graphicx}   
\usepackage{amsmath}    
\usepackage{amssymb}    
\usepackage{hyperref}   
\usepackage{xcolor}     
\usepackage{physics}    
\usepackage{siunitx}    
\usepackage{bm}         

\usepackage{comment}
\usepackage{amsmath}
\usepackage{xurl}
\usepackage[T1]{fontenc}
\usepackage[utf8]{inputenc}

\usepackage{enumitem}

\begin{document}

\title{Ligand Mediated {Magnetic} Coupling Across Metamagnetic Transitions in CrPS\texorpdfstring{$_4$}{4}}

\author{Giuseppe Buccoliero}
\affiliation{European Synchrotron Radiation Facility, 71 Avenue des Martyrs, CS 40220, 38043 Grenoble Cedex 9, France}
\affiliation{Université Grenoble Alpes, CNRS, Grenoble INP, Institut NEEL, 38000 Grenoble, France}

\author{Rachel Nickel}
\affiliation{European Synchrotron Radiation Facility, 71 Avenue des Martyrs, CS 40220, 38043 Grenoble Cedex 9, France}

\author{Roberto Sant}
\affiliation{Dipartimento di Fisica, Politecnico di Milano, Piazza Leonardo da Vinci 32, 20133 Milano, Italy}

\author{Marli dos Reis Cantarino}
\affiliation{European Synchrotron Radiation Facility, 71 Avenue des Martyrs, CS 40220, 38043 Grenoble Cedex 9, France}

\author{Andrei Rogalev}
\affiliation{European Synchrotron Radiation Facility, 71 Avenue des Martyrs, CS 40220, 38043 Grenoble Cedex 9, France}

\author{Nathan J. Yutronkie}
\affiliation{European Synchrotron Radiation Facility, 71 Avenue des Martyrs, CS 40220, 38043 Grenoble Cedex 9, France}

\author{Tristan Riccardi}
\affiliation{Université Grenoble Alpes, CNRS, Grenoble INP, Institut NEEL, 38000 Grenoble, France}

\author{Daniel A. Chaney}
\affiliation{European Synchrotron Radiation Facility, 71 Avenue des Martyrs, CS 40220, 38043 Grenoble Cedex 9, France}

\author{Kurt Kummer}
\affiliation{European Synchrotron Radiation Facility, 71 Avenue des Martyrs, CS 40220, 38043 Grenoble Cedex 9, France}

\author{Johann Coraux}
\affiliation{Université Grenoble Alpes, CNRS, Grenoble INP, Institut NEEL, 38000 Grenoble, France}

\author{Nicholas B. Brookes}
\affiliation{European Synchrotron Radiation Facility, 71 Avenue des Martyrs, CS 40220, 38043 Grenoble Cedex 9, France}

\date{\today}

\begin{abstract}
Chromium thiophosphate (CrPS$_4$) is a long-known material: a layered semiconducting antiferromagnet. Its recently discovered gate-tunable metamagnetic phase transitions, the remarkable positive and oscillating magnetoresistance as a tunnel barrier, and its Fano-resonance luminescence, elusive among the multitude of Cr$^{3+}$ compounds, call for revisiting the understanding of its electronic structure, especially regarding how it relates to magnetic order. Here, we employ X-ray magnetic circular dichroism, implemented in both absorption and resonant inelastic X-ray spectroscopies, together with quantum many-body calculations, to unveil {the role of metal--ligand covalency in mediating the metamagnetic transitions in CrPS$_4$, using crystal-field and charge-transfer excitations as fingerprints of the evolving magnetic order}. We reveal the role of extended superexchange paths involving P and S atoms, coupling interactions between the Cr spins across the different magnetic phases---antiferromagnetic, canted, and ferromagnetic. Our results elucidate the electronic states involved in these phases and provide prescriptions for engineering the metamagnetic phase diagram of CrPS$_4$.

\end{abstract}

\maketitle

\section{Introduction}
Chromium(III)-based inorganic semiconductors, e.g. garnets, spinels or sapphire, have long been known for their rich spin- and photo-physics. At least to a large extent, these rely on electronic ($d$--$d$) transitions involving Cr’s $d$ molecular orbitals, influenced by spin–orbit interaction and the host lattice, including structural distortions from a perfect octahedral local environment and coupling to phonons \cite{wood,Nelson1965,Castelli1975,Derkosch1977,Mikenda1981,Henderson1988,Adachi2020}. Many of the recently isolated few-atom-thick, magnetically ordered crystals (the so-called two-dimensional magnets) also rely on Cr$^{3+}$-based lattices \cite{gong2017discovery,CrI,Lee2017,Lee2021,Gu2023,samal2021two}, reflecting the same underlying physics. The parent layered compounds, from which they are exfoliated owing to the weak interlayer van der Waals bonding, have also been known for decades—chromium (chalcogenide) halides or tetrel chalcogenides \cite{Tsubokawa1960,Davis1964,Dillon1966,Christensen1974,Louisy1978,Beck1990,Goeser1990,Carteaux1995}—and their complex optoelectronic properties are currently under intense investigation, with critical insights gained in the past few years with the help of high-resolution spectroscopies \cite{CrIxmcd,occhialini2025spin,He2025,de2026orbital,bedoya2021intrinsic}.

Among these, chromium thiophosphate (CrPS$_4$) is an indirect bandgap semiconductor emitting light in the near-infrared and first synthesized in the late 1970s \cite{Louisy1978}. Below 38~K, successive ferromagnetic layers alternate, thereby forming an antiferromagnet \cite{zhuang2016density,adv}. This relative order is altered under the influence of an external magnetic field, via metamagnetic phase transitions (a spin-flop and a spin-flip one) \cite{adv}. Unlike most van der Waals magnets so far, CrPS$_4$ features large electronic bandwidth, such that thin flakes can serve as the conducting channel of a transistor, and the field values of its metamagnetic transitions can be tuned electrically \cite{Wu2023}. This way, strongly tunable magneto-conductance was demonstrated \cite{Wu2023}. Tunnel junctions with a CrPS$_4$ barrier moreover revealed very singular magnetoresistance, positive and oscillating \cite{Cheng2024,shi2024magnetoresistance,Lin2025}, related to the mediation of transport by in-gap states \cite{shi2024magnetoresistance,Lin2025}, also prompting the role of spin-dependent $d$--$d$ transitions and ligand effects \cite{Cheng2024}. In this context—and given the debated involvement of spin-forbidden transitions in the remarkable Fano resonances observed in CrPS$_4$ \cite{Gu2019,Riesner2022}—it becomes desirable to look beyond the standard band-structure picture. That picture places the Cr $e_g$ and $t_{2g}$ levels a few electronvolts apart within a $\sim 10$~eV S-P charge-transfer gap \cite{Ohno1989}, but misses the role of covalency, multiplet effects, and ligand-mediated pathways revealed by recent experiments, as well as their relation to the evolution of magnetic order. 

Here, we go beyond this picture: by using X-ray magnetic circular dichroism (XMCD)~\cite{xmcd}, resonant inelastic X-ray scattering (RIXS)~\cite{ament2011rmp}, and its dichroic variant (RIXS–MCD)~\cite{miyawaki2017dzyaloshinskii}, together with quantum many-body calculations~\cite{haverkort,quanty}, we analyze the $d$--$d$ and charge-transfer (Cr–ligand) transitions. Notably, RIXS--MCD is an element-specific probe of electronic excitations in magnetic materials~\cite{rixsmcd}; when performed in high magnetic fields, it uniquely allows the measurement and identification of those electronic states participating in the sequence of magnetic orders developing as a function of the magnetic field~\cite{magnuson2006large}.

Specifically, we unveil (\textit{i}) extended superexchange paths, involving both S and P and coupling ferromagnetically spins localized on Cr ions, (\textit{ii}) the anisotropy details of the Cr–ligand hybrid orbitals, and (\textit{iii}) how the Cr 3\emph{d} manifold and the hybridized excited states respond to and track the metamagnetic transitions. Our findings clarify the role of metal–ligand covalency in driving the metamagnetic response of CrPS$_4$, and should guide future efforts toward engineering magnetic order via ligand coordination effects.

\begin{figure}[t]
    \centering
    \includegraphics[width=\columnwidth]{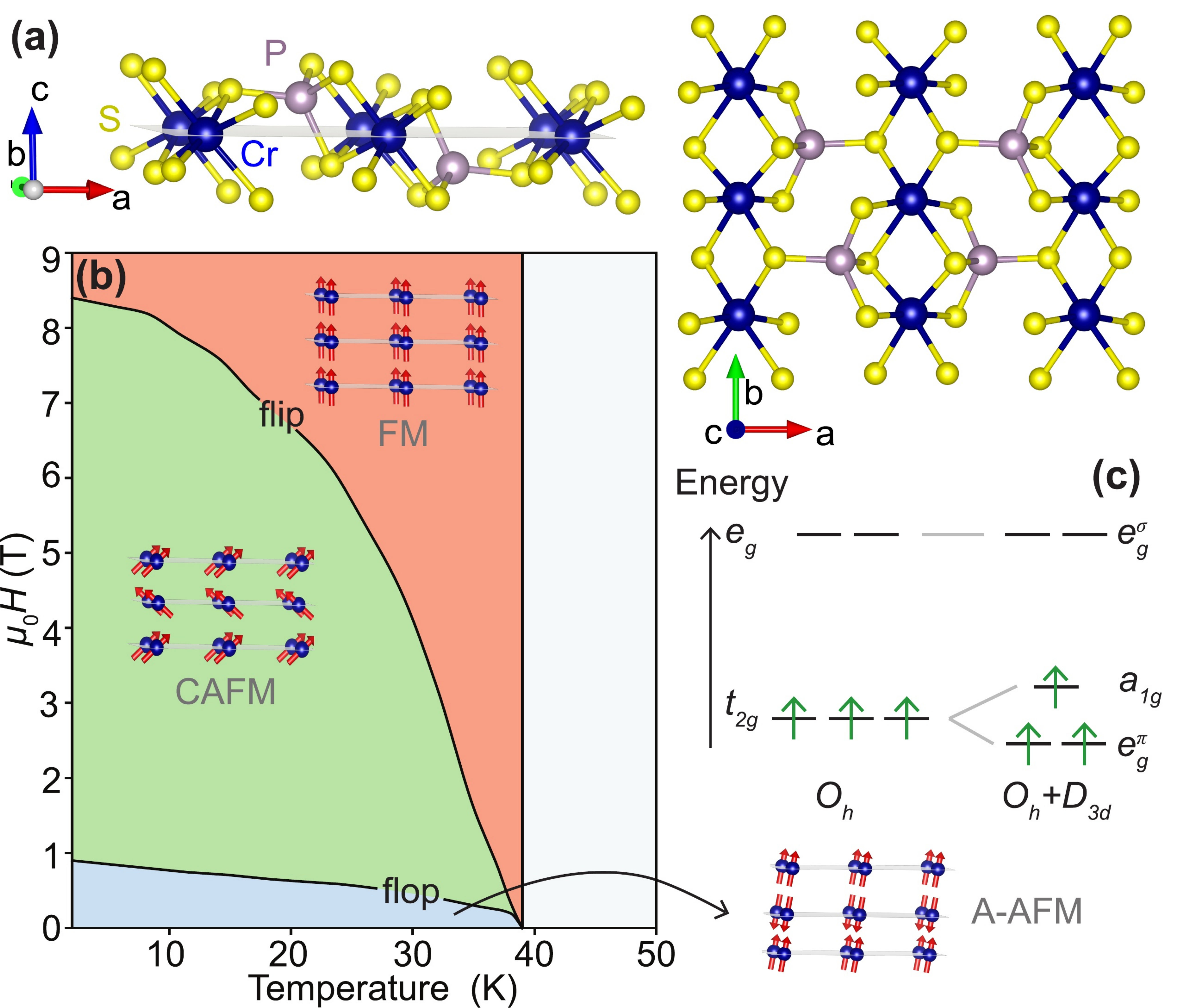}
    \caption{
    \textbf{Metamagnetic phases of CrPS$_4$.}
    (\textbf{a}) Side and top views of the crystal structure of CrPS$_4$, in which Cr atoms are octahedrally coordinated by S atoms.
    (\textbf{b}) Magnetic phase diagram of CrPS$_4$ (out of plane $\mu_0 H$ vs.\ $T$) reconstructed from magnetometry measurements.
    (\textbf{c}) Crystal-field splitting of Cr$^{3+}$ 3d states in octahedral ($O_h$) and trigonal-distorted ($O_h + D_{3d}$) cluster geometry, with the corresponding high-spin configuration.
    }
    \label{fig:fig1}
\end{figure}

\section{Methods}

\subsection{Experimental XMCD and RIXS-MCD Setups}
\input{text/sample_prep}

\subsection{Theoretical and Computational Framework}
\input{text/computation_methods}

\section{Results}
\subsection{Magnetic Structure and Field Induced Transitions}
\input{text/xmcd}

\begin{figure*}[t!]
  \centering
  \includegraphics[width=0.99\textwidth]{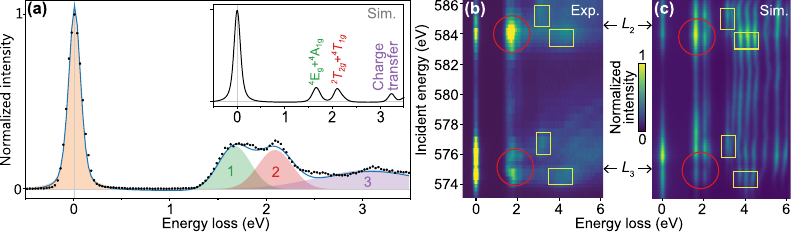}
\caption{\textbf{RIXS measurements and simulations at the Cr $L_{3,2}$ edges.}
(\textbf{a}) RIXS spectrum measured at the Cr $L_3$ incident energy (peak~A, 576.2\,eV) at 0\,T and 5\,K in NI geometry, overlaid with the corresponding fitting components. 
The labels identify the individual contributions associated with excitations labeled as peaks~1–3. 
\textit{Inset:} Simulated RIXS spectrum at the same incident energy, temperature, magnetic field, and geometry. 
(\textbf{b},~\textbf{c}) RIXS energy-loss maps, obtained as averages of spectra acquired with right- and left-circularly polarized light, across the Cr $L_{3,2}$ edges: (\textbf{b}) experimental and (\textbf{c}) simulated (0\,T, 5\,K, NI). 
The maps highlight the crystal-field excitations (red circles) and the charge-transfer excitations (yellow squares).}

  \label{fig:rixs}
\end{figure*}

\subsection{Field-Driven Electronic Excitations}

\input{text/rixs}

\section{Discussion}

\input{text/theory}

\begin{table*}[t]
\centering
\caption{Parameters used in the atomic multiplet calculations for Cr $L_{3,2}$-edge XMCD and RIXS--MCD simulations. All values are expressed in eV.}

\label{tab:params}

\small
\renewcommand{\arraystretch}{1.25}

\begin{tabular*}{\textwidth}{@{\extracolsep{\fill}}lccccccccccccc}
\hline\hline
& \multicolumn{4}{c}{$\mathcal{H}_{\mathrm{atomic}}$}
& \multicolumn{2}{c}{$\mathcal{H}_{\mathrm{CF}}$}
& \multicolumn{4}{c}{$\mathcal{H}_{\mathrm{LMCT}}$}
& $\mathcal{H}_{\mathrm{lig}}$
& \multicolumn{2}{c}{$\mathcal{H}_{\mathrm{mag}}$} \\
\cline{2-5} \cline{6-7} \cline{8-11} \cline{12-12} \cline{13-14}

& $U_{dd}$ & $U_{pd}$ & $\zeta_{3d}^{\mathrm{init}}$ & $\zeta_{3d}^{\mathrm{fin}}$
& $10Dq$ & $\tau$
& $\Delta$ & $V_\sigma$ & $V_\pi$ & $V_{a_{1g}}$
& $10Dq_{\mathrm{lig}}$
& $g_s\mu_B\mu_0 H_E$ & $\mu_B\mu_0 H_{\mathrm{ext}}$ \\
\hline

& 3.80 & 5.20 & 0.035 & 0.047
& 1.45 & 0.30
& 0.80 & 1.70 & 1.20 & 1.05
& 0.40
& \begin{tabular}{c}
$-1.0\times10^{-4}$ \\
$3.5\times10^{-2}$
\end{tabular}
& \begin{tabular}{c}
$0$ \\
$5.2\times10^{-4}$
\end{tabular}
\\[3pt]

\hline\hline
\end{tabular*}
\end{table*}

\section{Conclusion}

This work bridges microscopic electronic excitations with macroscopic magnetic behavior in CrPS$_4$ through advanced spectroscopies combined with atomic multiplet calculations. By combining X-ray magnetic circular dichroism with resonant inelastic X-ray scattering, we accessed element- and orbital-resolved magnetic responses and revealed how specific crystal-field and charge-transfer excitations evolve across the metamagnetic transitions. Abrupt changes in the magnetic dichroism directly track spin reorientation and the emergence of ferromagnetic order.

Quantum many-body simulations allowed us to extract key microscopic parameters, including spin and orbital moments, crystal-field splittings, and exchange-field strengths, yielding a coherent description of the evolution of the magnetic ordering. In particular, the emergence of magnetic dichroism at the spin-flop transition is consistent with a vanishing molecular exchange field in the AFM phase, leaving the spin-state manifolds nearly degenerate within $k_{\mathrm{B}}T$. The spin-flop transition generates an exchange-induced splitting of these manifolds, which leads to a finite dichroic response.

Ligand $K$-edge XMCD and Cr $L_{2,3}$ RIXS--MCD energy maps across the different metamagnetic phases further unveil the participation of sulfur and phosphorus in extended exchange pathways, modulating both anisotropy and spectral line shapes in the Cr$^{3+}$ spectra. Together, our results establish a broader principle: field-driven transitions in van der Waals magnets arise from a subtle interplay between hybridization, spin–orbit coupling, and magnetic exchange. Crucially, the ability to tune critical fields, such as the spin-flop and spin-flip thresholds, through ligand substitution offers a promising route to engineer 2D van der Waals magnets with tailored spin configurations and magnetoresistive responses. This tunability opens pathways toward functional magnetic architectures in vdW materials, wherein the hybridization between magnetic ions and their ligands, together with the resulting anisotropy and exchange constants, can be quantitatively analyzed using magnetic spectroscopies.

\vspace{-0.3em}
\section*{Conflict of Interest}
\vspace{-0.3em}
The authors declare no conflict of interest.
\vspace{0.9em}

\section*{Data Availability Statement}
The raw data supporting the findings of this study are available under the following DOIs:
\url{https://doi.org/10.15151/ESRF-ES-2159204286},
\url{https://doi.org/10.15151/ESRF-ES-2039934344}, and
\url{https://doi.org/10.15151/ESRF-ES-2186230392}.
The \textsc{Quanty} and LMTO scripts used for the simulations are available from the corresponding author upon reasonable request.

\begin{acknowledgments}
The authors thank the European Synchrotron Radiation Facility (ESRF) for providing beamtime under proposal numbers HC6092 and IHHC4112 at beamline ID32, and IHHC4179 at beamline ID12. We are grateful to the ID32 staff—particularly Dr.~Flora Yakhou-Harris, Dr.~Benedikt Eggert and Ms.~Celine Allais—for their technical support during the experiments. We also acknowledge Dr.~Laetitia Marty for facilitating the collaboration between Institut Néel (CNRS, Grenoble) and ESRF and Dr.~Luigi Paolasini for granting access to the ID28 beamline (ESRF, Grenoble) for diffuse scattering measurements. We further thank HQ Graphene for supplying the samples used in this study. This work was partially performed within the MUSA–Multilayered Urban Sustainability Action project, funded by the European Union–NextGenerationEU, under the National Recovery and Resilience Plan (NRRP) Mission~4 Component~2 Investment Line~1.5: Strenghtening of research structures and creation of R\&D, ``innovation ecosystems'' set up of ``territorial leaders'' in R\&D. This work is also supported by France 2030 government investment plan 7 managed by the French National Research Agency under grant reference PEPR SPIN - SPINMAT ANR-22-EXSP-0007.

\end{acknowledgments}

\bibliography{biblio}

\end{document}

%% file: text/sample_prep.tex
High-purity monocrystalline CrPS$_4$ flakes (99.995\%, purchased from HQ Graphene~\cite{hqgraphene}) with lateral dimensions on the millimeter scale were mounted on standard Omicron sample plates. Unless otherwise specified, samples were scotch-cleaved \textit{in situ} to expose fresh, clean surfaces to the beam. Further characterization details, including air stability, structural ordering, and domain configurations measured by X-ray diffuse scattering, are provided in the Supplemental Material~\cite{supplemental_material,air,girard2019new,agilent2014agilent,murayama2016crystallographic,wildes2023spin}.

XMCD measurements were performed at the European Synchrotron Radiation Facility (ESRF) on beamlines ID32~\cite{brookes2018beamline,ID32} and ID12~\cite{ID12,rogalev2013x}. Shown XMCD spectra were obtained as the difference between X-ray absorption spectra recorded with left- and right-circularly polarized light and signal is reported as a percentage of the average XAS edge jump. To suppress field-independent artifacts, the XMCD signal was antisymmetrized with respect to magnetic-field reversal ($\mathrm{XMCD}(H)=\tfrac{1}{2}\big(\mathrm{XMCD}(+H)-\mathrm{XMCD}(-H)\big)$). The uncertainty was estimated from the residual field-independent contribution ($\mathrm{error}(H)=\tfrac{1}{2}\big(\mathrm{XMCD}(+H)+\mathrm{XMCD}(-H)\big)$). 
 At ID32, experiments targeted the Cr L$_{3,2}$ absorption edges using circularly polarized soft X-rays at the High Field Magnet (HFM) endstation with a base pressure below $5 \times 10^{-10}$\,mbar. The incident beam exhibited nearly 100\% circular polarization, and the monochromator provided a resolving power better than 5000. At ID12, XMCD was carried out at the S and P $K$-edges using hard X-rays with an effective degree of circular polarization of approximately 21\% and 77\%, respectively. The XMCD data were corrected accordingly. The pressure was maintained below $1 \times 10^{-5}$\,mbar. Depending on the absorption edge and emission line, the energy resolution (FWHM) ranged between 160--200\,meV under standard operating conditions. All measurements were performed at selected temperatures under magnetic fields ranging from 0 to 9\,T, applied parallel to the incident beam. Both normal-incidence (NI, 80--90$^\circ$ with respect to the sample surface) and grazing-incidence (GI, 10--20$^\circ$) geometries were employed. {Specifically, in NI geometry, the incident beam and applied magnetic field are directed along the $c$ axis, the van der Waals stacking direction and easy axis of the metamagnetic transitions. In GI geometry, the beam and field lie nearly within the $ab$ plane, providing sensitivity to the in-plane orbital character.} X-ray absorption at ID32 was recorded in total electron yield mode and normalized to the incident flux. At ID12, total fluorescence yield mode was used, and self-absorption corrections were applied. XMCD signals are reported as a percentage of the average XAS edge jump. At the ligand K edges, the small XMCD signal required extensive averaging: each spectrum corresponds to the mean of 60 scans acquired while alternating both magnetic-field polarity and photon helicity. 

RIXS–MCD measurements were also conducted at ID32 using the dedicated RIXS spectrometer integrated into the HFM endstation, enabling acquisitions under identical low-temperature, high-field, and sample-geometry conditions as for XMCD. The spectrometer operates at a fixed scattering angle of 90$^\circ$. Emitted light was detected with a CMOS detector operated in total photon-integration mode. The total energy resolution was better than 250\,meV at the Cr L$_{3,2}$ edges, sufficient to resolve intra-atomic $d$–$d$ transitions and charge-transfer excitations, though not phonons or low-energy magnons. Although momentum resolution is fixed due to the geometry, the high photon throughput allowed efficient mapping of RIXS spectra across the Cr L$_{3,2}$ absorption edges. Beamline flux stability and polarization were routinely monitored to ensure reproducibility across magnetic-field cycles.

%% file: text/computation_methods.tex
Simulations of XMCD at the Cr $L_{3,2}$ edges and of RIXS–MCD spectra and energy maps were performed using the Anderson impurity model, as implemented in the quantum many-body package \textsc{Quanty}~\cite{haverkort,quanty}. The AMC employ a fully parameterized Hamiltonian for a single Cr$^{3+}$ cluster, where ligand-field effects are explicitly included through charge-transfer and hybridization terms {up to 
the $d^4\underline{L}$ configuration (single ligand hole)}. The local symmetry considered for the cluster is octahedral ($O_h$) with a trigonal distortion ($D_{3d}$), described by a distortion parameter corresponding to a compression along the threefold axis of the octahedron ($\tau>0$). The $10Dq$ crystal-field splitting and the trigonal distortion lift the degeneracy of the Cr~3$d$ levels as illustrated in Fig.~1(d):

\begin{equation}
\begin{array}{@{}l@{}}
E_{e_g^{\pi}} = -0.4 \times 10Dq - \dfrac{1}{3}\tau, \qquad
E_{a_{1g}} = -0.4 \times 10Dq + \dfrac{2}{3}\tau, \\
E_{e_g^{\sigma}} = +0.6 \times 10Dq~.
\end{array}
\end{equation}

The AMC are based on the following multi-parameter Hamiltonian:
\begin{equation}
\mathcal{H}_{\mathrm{tot}}=
\mathcal{H}_{\mathrm{atomic}}+
\mathcal{H}_{\mathrm{CF}}+
\mathcal{H}_{\mathrm{LMCT}}+
\mathcal{H}_{\mathrm{ligand}}+
\mathcal{H}_{\mathrm{mag}}\, ,
\end{equation}
whose optimized parameters, used consistently for both XMCD and RIXS simulations, are reported in Table~\ref{tab:params}.
Importantly, $\mathcal{H}_{\mathrm{mag}}$ includes both a Zeeman term describing the applied external magnetic field
(i.e. $\mu_B\,\mu_0 \mathbf{H}_{\mathrm{ext}}\!\cdot(\mathbf{L}+2\mathbf{S})$)
and an effective internal molecular mean field
(i.e. $g_s\mu_B\,\mu_0 \mathbf{H}_{E}\!\cdot \mathbf{S}$),
accounting for the exchange interaction.

S and P $K$-edge XMCD simulations were also performed using a fully relativistic linear muffin–tin orbital (LMTO) framework~\cite{andersen1975linear}. Calculations employed the spin-polarized, relativistic implementation of LMTO with a Perdew–Wang exchange–correlation potential~\cite{perdew1992accurate}, using experimental crystal-structure parameters~\cite{zhuang2016density}. Further details on the models are provided in the Supplemental Material~\cite{supplemental_material}.

%% file: text/xmcd.tex
CrPS$_4$ crystallizes in the monoclinic $C2$ structure: within each layer, edge-sharing CrS$_6$ octahedra are bridged along the $a$ axis by isolated PS$_4$ tetrahedra, as illustrated in Fig.~1(a). Along $b$, the octahedra connect into quasi-one-dimensional chains, conferring pronounced in-plane structural anisotropy~\cite{houmes2024highly}. The charge-neutral layers stack along $c$ via weak van der Waals forces, which underpin the material’s two-dimensional magnetic behavior~\cite{diehl1977crystal}.
Figure~1(b) presents the magnetic phase diagram of CrPS$_4$ under an out-of-plane magnetic field, along the $c$ axis, reconstructed from our magnetometry measurements and previously reported data~\cite{fkas2025direct,adv}. Below the N\'eel temperature ($T_N \approx 38$\,K), the material exhibits A-type antiferromagnetic (A--AFM) order: spins are ferromagnetically aligned within each layer, while adjacent layers are oppositely magnetized. Applying a sufficiently strong field ($\mu_0 H_{\mathrm{flop}}$) along $c$ induces a spin-flop transition, as commonly observed in layered antiferromagnets~\cite{basnet2021highly}. Upon crossing $\mu_0 H_{\mathrm{flop}}$, CrPS$_4$ enters a canted antiferromagnetic (CAFM) state, where the layer magnetizations no longer fully cancel, resulting in a finite net moment along $c$~\cite{adv}. At higher fields ($\mu_0 H_{\mathrm{flip}}$), all spins align with the field direction and the system reaches a fully polarized ferromagnetic (FM) state. A peculiarity of CrPS$_4$ is its exceptionally low critical fields. For comparison, MnPS$_3$ exhibits a spin-flop transition around 5\,T at 5\,K~\cite{MnPS3}, NiPS$_3$ around 6\,T under an in-plane field~\cite{SF_Ni}, and other MPS$_3$ compounds require even higher fields~\cite{basnet2021highly}. 

CrPS$_4$'s characteristic magnetic transitions and spin ordering mostly originate from the electronic structure of its Cr$^{3+}$ ions: we model the local Cr$^{3+}$ environment as an octahedral crystal field with a slight trigonal distortion, which splits the five Cr 3$d$ orbitals into low-lying $e_g^{\pi}$ (doublet) and $a_{1g}$ (singlet) levels, and higher-lying $e_g^{\sigma}$ levels, as illustrated in Fig.~1(c). In a 3$d^3$ configuration, a high-spin ground state ($S = 3/2$) is typically realized, corresponding to three unpaired electrons occupying the half-filled $t_{2g}$-like manifold (term ${}^4A_2$)~\cite{achinuq}. { XMCD measurements, combined with multiplet calculations, allows us to directly quantify the Cr$^{3+}$ spin and orbital magnetic moments, providing direct microscopic insight into the electronic ground state.}
\begin{figure}[b]
  \centering
  \includegraphics[width=\columnwidth]{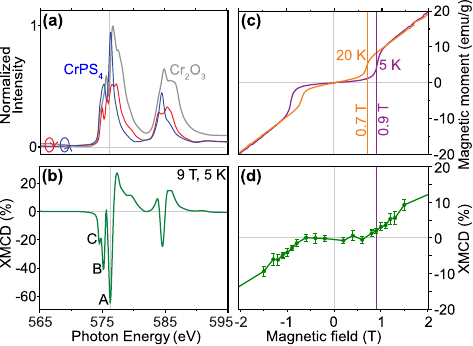}
  \caption{\textbf{Magnetic response of CrPS$_4$ across the metamagnetic transition via XMCD.}
  (\textbf{a}) Cr $L_{3,2}$-edge XAS spectrum of CrPS$_4$ at 5\,K, overlaid with a reference Cr$_2$O$_3$ spectrum, emphasizing absolute energy positions and the chemical shift. 
  (\textbf{b}) XMCD spectrum of CrPS$_4$ at 5\,K under a 9\,T magnetic field (NI geometry), normalized to the maximum XAS amplitude. The XMCD is obtained as the difference between XAS spectra recorded with opposite photon helicities.
  (\textbf{c}) SQUID magnetometry at 5\,K and 20\,K for out-of-plane fields, with vertical lines indicating the spin-flop transition. 
  (\textbf{d}) Cr$^{3+}$ element-specific magnetization at 5\,K (NI geometry), extracted from the XMCD intensity at peak~A (576.2\,eV). Error bars correspond to the residual field-independent contribution.}
  \label{fig:xmcd}
\end{figure}
\begin{figure*}[t]
  \centering
  \includegraphics[width=0.95\textwidth]{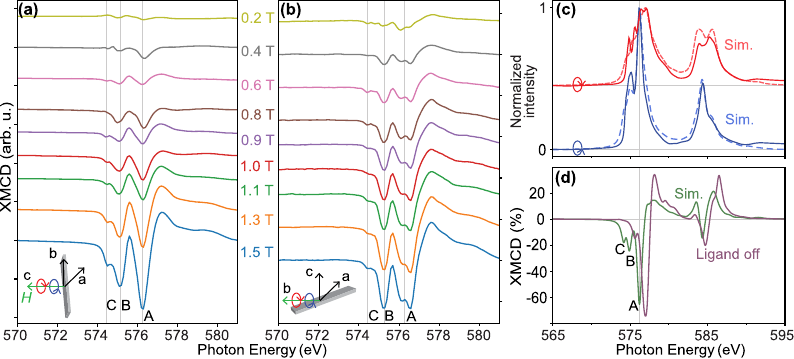}
  \caption{\textbf{Ligand effects in CrPS$_4$.}
  (\textbf{a},~\textbf{b}) Field-dependent XMCD at the Cr $L_3$ edge of CrPS$_4$ (5\,K) across the spin-flop transition ($\mu_0 H_{\mathrm{flop}} \approx 0.9$\,T at 5\,K for $\mu_0H \parallel c$), measured in NI (a) and GI (b) geometries.
  {\textbf{(c)} Simulated Cr $L_{3,2}$-edge XAS spectra (9\,T, 5\,K --- dashed lines) are overlaid with experimental Cr $L_{3,2}$-edge spectra recorded with right- and left-circularly polarized light (9\,T, 5\,K). Spectra are vertically shifted for clarity.} 
  (\textbf{d}) Simulated XMCD spectrum at 5\,K in a 9\,T field (NI geometry). 
  The green curve includes ligand-field and charge-transfer effects, whereas the purple curve corresponds to the simulated XMCD obtained without ligand parameters.}
  \label{fig:rixs_panel}
\end{figure*}
XMCD measurements were carried out as a function of the applied magnetic field along the beam direction at the Cr $L_{3,2}$ edges, in both normal-incidence (NI) and grazing-incidence (GI) geometries. Figure~2(a) shows the average Cr $L_{3,2}$ XAS spectrum together with the corresponding XMCD measured at 5\,K in a 9\,T field [Fig.~2(b)].

A large dichroic signal, approximately 60\% of the XAS intensity, is observed, consistent with a saturated ferromagnetic state and comparable to XMCD intensity in other vdW ferromagnets~\cite{bedoya2021intrinsic}. The XMCD spectrum exhibits three distinct features at the $L_3$ edge: a main peak at $\sim$576.2\,eV (A) and two weaker pre-edge peaks at $\sim$575.0\,eV (B) and $\sim$574.6\,eV (C). For comparison, the XAS of Cr$_2$O$_3$ (a reference for Cr$^{3+}$) is plotted in Fig.~2(a). The CrPS$_4$ spectrum is chemically shifted to lower energy by $\sim$0.5\,eV, indicating a slightly lower effective Cr oxidation state and higher covalency. This is supported by XMCD simulations, which yield a higher 3$d$ electron count for CrPS$_4$ (estimated $N_{3d} = 3.44$). Sum-rule analysis~\cite{thole1992x} reveals an orbital contribution $\langle L_z \rangle < 0.07\,\mu_\mathrm{B}$, confirming that the Cr$^{3+}$ orbital moment is nearly quenched~\cite{chen2026probing}. This provides direct microscopic evidence of weak spin–orbit coupling and explains the intrinsically low magnetic anisotropy in CrPS$_4$, consistent with its unusually low transition fields compared to MPS$_3$ compounds~\cite{kim,calder2020prb}.

By comparing magnetometry and XMCD, it is possible to directly connect the macroscopic magnetic properties of CrPS$_4$ with its microscopic electronic structure. At 5\,K, SQUID magnetometry [Fig.~2(c)] clearly shows the spin-flop transition at $0.9 \pm 0.15$\,T, marked by a sharp enhancement in the magnetic moment. $\mu_0 H_{\mathrm{flop}}$ is temperature dependent, shifting to $0.7 \pm 0.25$\,T at 20\,K. This reduction arises from thermally activated excitations as well as from the temperature dependence of the anisotropy and exchange constants~\cite{seo2024probing}. The transition is relatively broad at 5\,K ($\sim$0.3\,T) and becomes broader at higher $T$ ($\sim$0.5\,T). Fig.~2(d) shows the element-specific magnetization curve for Cr$^{3+}$ at 5\,K, measured from the XMCD intensity at the peak A photon energy as a function of field. The curve exhibits a sharp rise at $\sim$0.9\,T ($\mu_0 H_{\mathrm{flop}}$), followed by a linear increase until saturation, closely mirroring the bulk magnetization measured by SQUID [Fig.~2(c)]. Remarkably, the XMCD signal becomes finite immediately after the spin-flop transition: from its intensity (proportional to the Cr moment projected along the beam), we estimate a spin-canting angle of $\sim$85$^\circ$ from the $c$ axis just above the transition. Measurements at 20\,K show similar behavior, with critical fields shifting to lower values (see Supplemental Material~\cite{supplemental_material}), further validating the correspondence.

\subsection{Ligand Effects and Exchange Mechanism}

XMCD spectra at 5\,K under various magnetic fields, collected in both normal and grazing-incidence geometries, provide further microscopic insight. In NI geometry [Fig.~3(a)], the XMCD evolves through the spin-flop transition: at low fields, the signal is nearly zero and featureless—consistent with the antiparallel spin configuration in the A-AFM phase—whereas above $\mu_0 H_{\mathrm{flop}}$ ($\sim$0.9\,T), the characteristic three-peak XMCD spectrum emerges.

{ Above $H_\mathrm{flop}$, the spectral line shape remains unchanged and the signal increases monotonically toward saturation, as measured up to 9\,T (cf.\ Fig.~2(b)).} Around the transition field ($\sim$0.8–0.9\,T), the most intense XMCD features (peaks A and B) start to develop, although the full fine structure (peak C) is not yet fully resolved. In contrast, the GI spectra [Fig.~3(b)] remain unchanged in shape across the same field range. Both NI and GI measurements eventually reach similar maximum XMCD intensities at saturation; however, while the peak positions coincide, the relative intensities differ. In particular, in GI geometry, peak A appears broader and nearly split, peak B has comparable intensity with A, and peak C is almost suppressed.

These differences reflect the anisotropy, arising from the orbital character of the unoccupied states and the local ligand-field environment~\cite{thole1992x}. Comparing with the atomic multiplet calculations (AMC) in the two different geometries (see Supplemental Material~\cite{supplemental_material}), the asymmetry between peaks B and C reflects the spatial character of the metal--ligand hybrid orbitals: $e_g^{\sigma}$ and $a_{1g}$ hybridizations involve out-of-plane $\sigma$ bonding, while $e_g^{\pi}$ corresponds to in-plane $\pi$ bonding. Consequently, the GI geometry probes $\pi$-character states more strongly, whereas NI gives enhanced sensitivity to $\sigma$-character transitions, leading to geometry-dependent changes in the pre-edge spectral features that are highly sensitive to the different ligand hybridization potentials included in the simulations~\cite{Stohr1999JMMM,haverkort2012multiplet}. 

{This geometry dependence is more pronounced than in the related vdW ferromagnet Cr$_2$Ge$_2$Te$_6$, where the NI--GI anisotropy is governed primarily by the trigonal crystal-field splitting~\cite{suzuki2022magnetic}. In CrPS$_4$, the trigonal distortion similarly contributes, but the stronger metal--ligand covalency, reflected in a lower charge-transfer energy, drives a pronounced orbital anisotropy through the symmetry-specific hybridization potentials.} 

Figures~3(c-d) displays simulated Cr $L_{3,2}$-edge XAS and XMCD spectra in the FM phase obtained from AMC. The simulated curves include full ligand-field and charge-transfer terms (see Methods) and successfully reproduce the experimental XAS with opposite circularly polarized light (overlaid curves), as well as the XMCD fine structure, including peaks B and C in Fig.~3(d). Conversely, when all ligand contributions are turned off [purple curve in Fig.~3(d)], peaks B and C vanish and the overall line shape becomes markedly different, shifting to higher energy. These results highlight that the microscopic signature of the spin-flop transition, revealed in the Cr $L$-edge XMCD spectra by the emergence of the characteristic pre-edge doublet (peaks B and C), ultimately stems from ligand-field effects. 

To gain deeper insight into the role of the ligand environment, we investigated the XMCD at the sulfur and phosphorus $K$ edges. Figure~4(a-b) shows the XAS and XMCD spectra at the S $K$-edge of CrPS$_4$, measured in NI at 5\,K in an 8\,T field. A finite XMCD signal of $0.75 \pm 0.05\,\%$ with respect to the edge jump is detected at the S $K$-edge. This XMCD signal indicates that the ligands carry an induced spin polarization, confirming that the magnetic interactions in CrPS$_4$ are transmitted via exchange through the sulfur atoms. This M–S exchange mechanism is crucial in tuning these field-induced transitions in the related MPS$_3$ compounds (e.g. MnPS$_3$~\cite{ga2025interlayer}), and element-specific magnetization curves at the S $K$-edge (Supplemental Material~\cite{supplemental_material}) show a sharp increase for the sulfur XMCD signal at 0.9\,T as observed in SQUID macroscopic magnetization curve. More surprisingly, we also observe an XMCD signal at the P $K$-edge as shown in Fig.~4(c-d), of comparable magnitude ($0.60 \pm 0.04$\,\% of the edge jump) under the same temperature, field, and geometry conditions. The detection of such a signal, despite the typically non-magnetic nature of phosphorus, is highly unusual~\cite{guillou2015electronic} and implies that the exchange mechanism in the FM phase extends to the P atoms, revealing a Cr–S–P–S–Cr superexchange pathway.

\begin{figure}[b]
  \centering
  \includegraphics[width=\columnwidth]{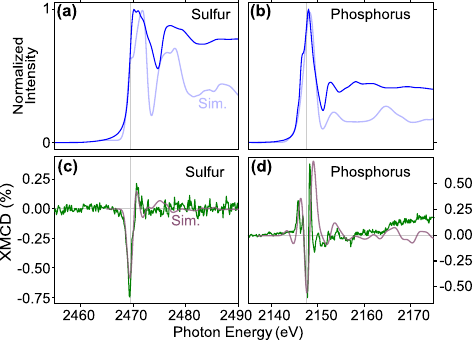}
\caption{\textbf{Induced magnetic moment on S and P.}
(\textbf{a},\textbf{b}) XAS spectra of CrPS$_4$ measured at 5\,K in an applied field of 8\,T (normal incidence) at the S $K$-edge (a) and P $K$-edge (b), overlaid with the corresponding LMTO-simulated XAS.
(\textbf{c},\textbf{d}) Corresponding XMCD spectra at the S $K$-edge (c) and P $K$-edge (d), obtained as the difference between right- and left-circularly polarized absorption, and compared with the LMTO-simulated XMCD.}

 \label{fig:rixs}
\end{figure}

Our linear muffin–tin orbital (LMTO) simulations predict that the net induced spin on the ligands amounts to $-0.11\,\mu_{\mathrm{B}}$ over the three non-equivalent S sites and $+0.035\,\mu_{\mathrm{B}}$ on P. The opposite signs reflect distinct hybridization paths: S 3$p$ states couple directly to the metal $d$ manifold, producing an antiparallel $p$--$d$ spin polarization and a less structured XMCD profile [Fig.~4(c)], whereas the P 3$p$ spin states are polarized parallel to the Cr spins indirectly via covalent S--P $\sigma$ bonds within the ($\mathrm{P_2S_6}$) units, giving rise to a more complex XMCD line shape [Fig.~4(d)]. Phosphorus tetrahedral coordination and strong covalent bonding with surrounding S atoms provide an efficient bridge for spin-polarized charge transfer: this behavior is consistent with $K$-edge XMCD analyses~\cite{xmcd} and conforms to the Goodenough--Kanamori--Anderson rules for superexchange through non-magnetic bridges in insulating magnets~\cite{goodenough,kanamori1959superexchange,anderson1950antiferromagnetism}. {The participation of P--S bonds in modulating these exchange pathways is further corroborated by a concurrent Raman study on CrPS$_4$, which reveals anomalous spin--phonon coupling of the P--S rocking mode, another signature of its coupling to the Cr--S--Cr superexchange geometry~\cite{pandey2026spin}.}

%% file: text/rixs.tex
RIXS--MCD can reveal subtle magnetism-associated spectral features while disentangling overlapping electronic excitations, complementing the XMCD information~\cite{miyawaki2017dzyaloshinskii}. 

Figure~5(a) shows a representative Cr $L_3$-edge ($2p_{3/2}$) RIXS spectrum of CrPS$_4$ measured at 5\,K. Three distinct inelastic peaks (in addition to the elastic peak) are clearly resolved at 1.69\,eV (peak~1), 2.04\,eV (peak~2), and 3.15\,eV (peak~3) energy loss. A simulated RIXS spectrum [inset in Fig.~5(a)] reproduces these features and aids their interpretation. {Peaks~1 and~2 correspond to crystal-field excitations of the Cr$^{3+}$ 3$d$ levels. Peak~1 encompasses the spin-conserving $^4E_g$ and $^4A_{1g}$ components of the trigonally-split $^4T_{2g}$ multiplet in $D_{3d}$ symmetry, together with the spin-flip $^2E_g$ and $^2T_{1g}$ transitions. Peak~2 corresponds the $^4T_{1g}$ and $^2T_{2g}$ excitations. At the experimental resolution of $\sim$250\,meV, these groups of transitions are not individually resolved; their assignment is established through high-resolution RIXS measurements and AMC simulations (Sec.~S3~\cite{supplemental_material}). The energy of peak~1 is primarily determined by the octahedral crystal-field strength ($10Dq = 1.45$\,eV in the AMC) and by the degree of covalency, parameterized through the Slater integrals and the Coulomb interaction energies. The separation between peaks~1 and~2 reflects the combined effect of these Coulomb and exchange interactions together with the ligand hybridization potentials and the trigonal distortion governing the fine splitting.} Peak~3, appearing above 3\,eV, corresponds to the first charge-transfer (CT) excitation, with a charge-transfer energy $\Delta = 0.80$\,eV in the model—consistent with the onset of both the third experimental peak and the simulated CT feature. Indeed, in the CT multiplet framework, $\Delta$ controls the energy threshold for $3d^n \leftrightarrow 3d^{n+1}\underline{L}$ hybridization (where $\underline{L}$ denotes a ligand hole), rather than the absolute energy loss of the CT features, which are also influenced by multiplet effects and core-hole interactions~\cite{ament2011rmp}. Notably, a pronounced gap is observed between the elastic line and the first $d$--$d$ excitation. In other MPS$_3$ compounds, such as FePS$_3$, which share a similar electronic structure, this energy region hosts several non-dispersive excitations~\cite{wei2025spin}. In that case, the excitations have been attributed to strong spin--orbit coupling and orbital anisotropy~\cite{dhakal2024hybrid}—ingredients absent in CrPS$_4$, where $\langle L_z \rangle$ is almost quenched.

Figures~5(b) and~5(c) present the full RIXS intensity maps as a function of incident photon energy, spanning the Cr $L_3$ and $L_2$ absorption edges at a fixed momentum transfer $|\mathbf{q}| = 0.40$--$0.43$\,\AA$^{-1}$ for the experiment and the AMC calculations, respectively. The two crystal-field excitations (red circles) appear broad and intense, together with higher-energy groups of CT excitations identified at the $L_{3,2}$ edges (yellow squares), in close analogy to the RIXS energy maps reported for CrSBr, a layered magnetic semiconductor~\cite{poree2025resonant}. The incident-energy dependence reflects resonant enhancement at the absorption edges. Importantly, the energy positions of these loss features remain essentially unchanged across different magnetic phases, indicating that the magnetic transitions do not significantly alter the $d$--$d$ excitation energies or the CT gap (full spectra are shown in the Supplemental Material~\cite{supplemental_material}). However, their intensities vary strongly with both incident energy and polarization.

Figure~6(a) shows the RIXS--MCD intensity map measured in the fully saturated ferromagnetic phase (9\,T, 5\,K) in NI geometry. The $d$--$d$ excitations (1.7–2.2\,eV loss) and the CT features (loss $>$ 3\,eV) exhibit strong MCD: under left- and right-circular polarization, their intensities differ markedly, with opposite signs at the $L_3$ and $L_2$ edges. The presence of well-defined magnetic dichroism in the CT excitations demonstrates that ligand-hybridized states are strongly coupled to the magnetic transitions. Both the RIXS intensity and the dichroism are particularly enhanced at the $L_2$ edge. In contrast, the dichroism at the $L_3$ edge is broader and more complex, reflecting the richer multiplet structure observed in the XMCD $L_3$ edge. The enhanced response at the $L_2$ edge arises from the $j$-dependent effective scattering operators, which provide greater selectivity for the $2p_{1/2}\!\rightarrow\!3d$ channel, as well as from the shorter $2p_{1/2}$ core-hole lifetime. The latter limits intermediate-state relaxation (e.g. spin--lattice coupling and screening), preserving the circular-dichroic contrast. In comparison, the longer lifetime at the $L_3$ edge allows stronger relaxation effects, which tend to suppress the asymmetry~\cite{van1991strong,veenendaal2006prl}. A similar enhancement of $L_2$-edge dichroism has been reported in other transition-metal systems, such as Mn--Zn ferrite~\cite{magnuson2006large}.

To track the evolution of RIXS--MCD with magnetic order, we measured dichroic maps for each magnetic phase under identical geometry conditions. Figure~6(b) shows the map at 0\,T (5\,K, A--AFM state). The dichroic signal is largely quenched and, in the $d$--$d$ excitation region, reverses sign, consistent with the antiferromagnetic layer stacking producing no net moment along the beam direction.
\begin{figure} [!]
  \centering
  \includegraphics[width=81mm]{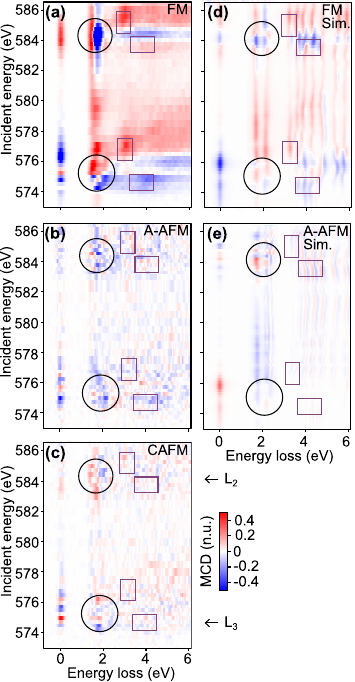}
\caption{\textbf{Cr $L_{3,2}$-edges RIXS-MCD energy maps across the metamagnetic transitions}
(\textbf{a--c}) RIXS–MCD energy maps, obtained as the difference between RIXS spectra acquired with right- and left-circularly polarized light and normalized to the global maximum MCD intensity among all maps, across the Cr $L_{3,2}$ edges at 5\,K and NI geometry: 
(\textbf{a}) 9\,T (FM phase); 
(\textbf{b}) 0\,T (A-AFM phase); 
(\textbf{c}) 1.1\,T (CAFM phase); 
(\textbf{d},~\textbf{e}) Simulated RIXS–MCD energy maps across the Cr $L_{3,2}$ edges at 5\,K in NI geometry, normalized to the global maximum MCD intensity: 
(\textbf{d}) FM phase (9\,T); 
(\textbf{e}) A-AFM model (0\,T; the signal intensity is amplified by a factor of $\times 10$ for better readability). 
}

  \label{fig:theory}
\end{figure}
After the spin-flop transition, the MCD rapidly emerges across the crystal-field excitations, weaker (about one order of magnitude less) but with the same sign as observed in the FM phase, reflecting the net magnetic moment in the CAFM state, as shown in Fig.~6(c).
 
Many-body simulations allow us to disentangle the various spectral contributions. By modeling XMCD and RIXS with the same parameter set, including the RIXS--MCD energy maps across the Cr $L_{3,2}$ edges, we impose tighter constraints on the Hamiltonian, an essential strategy for validating complex, multi-parameter models and identifying the underlying driving mechanisms. A key aspect of modeling the metamagnetic transition is the inclusion of both the external magnetic field ($\mu_0 H_{\mathrm{ext}}$) and an internal molecular exchange field ($\mu_0 H_{\mathrm{E}}$) in the Hamiltonian, which ultimately are responsible for the magnetic dichroism in the calculations. We track the evolution of the spectra by assigning distinct $\mu_0 H_{\mathrm{E}}$ values. {We stress that within our model, $\mu_0 H_{\mathrm{E}}$ is a phenomenological parameter that mimics the net effect of the exchange interactions on the local Cr$^{3+}$ site. Within the inherent limitations of a single-site cluster approach, setting $H_{\mathrm{E}}$ to a small value effectively captures the suppression of the net MCD signal arising from the near-cancellation of contributions from oppositely magnetized Cr sublattices in the 
A--AFM phase.}

Specifically, in the zero-field A--AFM phase (with antiparallel spin alignment along the $c$ axis), $\mu_0 H_{\mathrm{E}}$ is set small and negative along $c$, with a magnitude of $\mu_0 |H_{\mathrm{E}}| \approx 0.8\,\mathrm{T}$. This corresponds to an exchange energy of $4.63 \times 10^{-5}$\,eV per $\mu_{\mathrm{B}}$ (i.e., about $0.14$\,meV per Cr ion for a $3\,\mu_{\mathrm{B}}$ moment~\cite{codata}), which defines the energy barrier between the A--AFM and CAFM phases, in agreement with previous determinations of the exchange constant across this metamagnetic transition~\cite{seo2024probing}. When applied along $c$, $\mu_0 H_{\mathrm{ext}}$ competes with both $\mu_0 H_{\mathrm{E}}$ and the magnetic anisotropy, eventually driving the spin-flop transition once $H_{\mathrm{ext}} > H_{\mathrm{E}}$ in our AMC. Beyond this point, the effective molecular exchange energy saturates at $\sim 0.035$\,eV in the FM phase. Figure~6(d) shows the simulated RIXS--MCD intensity map for the FM phase in NI geometry. The strong dichroic response across the $3d$ excitations reproduces the experimental trends, with enhanced MCD intensity at the $L_2$ edge. A narrow energy-loss region exhibits a full MCD inversion at the $L_2$ edge, as observed experimentally. The AMC yield the expectation values of the spin and orbital moments in the FM phase: $\langle S_{z} \rangle \simeq 2.92\,\mu_{\mathrm{B}}$ (close to the nominal $S = 3/2$ value for high-spin Cr$^{3+}$ and consistent with LMTO estimates, $\langle S_{z} \rangle \approx 2.99\,\mu_{\mathrm{B}}$) and a quenched orbital moment $\langle L_{z} \rangle \simeq 0.04\,\mu_{\mathrm{B}}$, in good agreement with the XMCD sum-rule estimate ($\langle L_{z} \rangle < 0.07\,\mu_{\mathrm{B}}$). In the FM phase, the high-spin $S = 3/2$ state constitutes the ground state, well separated from the next-lowest excited state ($\Delta E = 35.8$\,meV), resulting in an almost full thermal occupation ($\sim$100\%) at 5\,K. As the effective exchange field is reduced, this splitting collapses. In the zero-field A--AFM phase, the $S=3/2$ manifold recovers a degenerate set of four spin-projection components, with extracted values $|\langle S_z\rangle| \approx 2.92\,\mu_{\mathrm{B}}$ and $|\langle S_z\rangle| \approx 0.94\,\mu_{\mathrm{B}}$. As a result, their thermal populations are comparable, leading to a suppression of the dichroic signal as shown in Fig.~6(e), where the signal is weak and shows a partially inverted sign, consistent with experiment. {This suppression is fully consistent with a two-sublattice picture: in the A--AFM phase, the MCD contributions from the two Cr sublattices with opposite spin orientations cancel, and the residual signal arises only from a small finite magnetic moment induced by the applied field.} Overall, the RIXS--MCD analysis shows how the Cr $3d$ excitations, both the crystal-field $d$--$d$ transitions and the charge-transfer features, respond to changes in magnetic order, highlighting the technique as a sensitive element- and site-specific probe for tracking magnetic phase evolution in van der Waals materials.

%% file: text/theory.tex
Our combined analysis provides a coherent picture of metamagnetism in CrPS$_4$ in terms of a strongly correlated, covalent Cr$^{3+}$ multiplet coupled to ligand-mediated superexchange. Multiple $d$--$d$ excitations with sizable RIXS--MCD signals have also been observed in CrI$_3$~\cite{ghosh2023magnetic}, where the $d$--$d$ energies remain essentially unchanged across $T_C \simeq 61$\,K, while a dichroic signal appears only below $T_C$ under an applied field of $\sim$0.4\,T. This behavior was attributed to a dominant local intra-atomic exchange field and a vanishingly small interatomic ferromagnetic superexchange contribution. We clarify and extend this picture for CrPS$_4$, consistently with our XMCD and RIXS results and modeling, in which the effective exchange $\mu_0 H_{\mathrm{E}}$ drives the magnetic dichroism of both the $d$--$d$ and CT excitations across the metamagnetic transitions. {In particular, if we consider a two-sublattice uniaxial approximation, the spin-flop field is proportional to both the interlayer exchange and the anisotropy field~\cite{formula_flop,seo2024probing,bogdanov2007spin}. In our framework, the anisotropy field arises self-consistently from the combined action of spin--orbit coupling and crystal-field symmetry. The unusually low critical fields in CrPS$_4$ then follow from both a relatively small interlayer exchange and weak anisotropy, consistent with the quenched orbital 
moment inferred from XMCD sum rules and with the inferred narrow energy barrier between the A--AFM and CAFM phases.} 
Our results are also consistent with optical studies on CrPS$_4$. Susilo \textit{et al.} identify two spin-allowed Cr$^{3+}$ $d$--$d$ transitions (T$_1$ and T$_2$) around 1.6--1.8\,eV, and a higher-energy feature of ligand-to-metal CT character~\cite{susilo2020band}. Our RIXS peaks at 1.69\,eV and 2.04\,eV fall naturally into this scheme, and the CT onset around 3.15\,eV is consistent with a covalent Mott-like Cr--S--P system. The large Coulomb interaction parameters used in our simulations ({$U_{dd} = 3.80$\,eV, the on-site $3d$--$3d$ repulsion, and $U_{pd} = 5.20$\,eV, the $2p$ core-hole--$3d$ interaction}) confirm this view of CrPS$_4$ as a strongly correlated covalent insulator.
Similar ligand-driven effects have been reported in other Cr-based van der Waals magnets. In CrI$_3$, angle-resolved photoemission and XMCD demonstrate the active role of ligands, with covalency contributing significantly to magnetic anisotropy~\cite{de2026orbital,CrIxmcd}. Multiplet simulations for chromium chlorides (i.e., CrCl$_3$, CrCl$_2$) highlight the dominant influence of charge transfer on XMCD line shapes, which show larger CT energies and weaker pre-edge structures~\cite{buccoliero2025situ}. {High-resolution RIXS on CrI$_3$ further reveals that spin-flip ($\Delta S = 1$) ${}^{4}A_{2g} \!\rightarrow\! {}^{2}E_g / {}^{2}T_{1g}$ transitions become visible below 2.5\,eV energy loss and carry strong dichroism when covalency and SOC are sufficiently large~\cite{occhialini2025spin}. The same physics governs CrPS$_4$, where the explicit trigonal distortion additionally resolves the $^2E_g$ from $^2T_{1g}$, as confirmed by our high-resolution RIXS and AMC simulations (Sec.~S3~\cite{supplemental_material}).}

The role of ligand chemistry in controlling exchange and anisotropy extends beyond Cr-based van der Waals magnets. In FePS$_3$, strong metal--ligand hybridization is required to explain its large out-of-plane anisotropy and sizable orbital moment~\cite{lee2023giant,wei2025spin}. Systematic studies on MPS$_3$ and their Se/Te analogs show that replacing S with heavier chalcogens tends to reduce anisotropy and exchange fields~\cite{basnet2021highly,yamagami2021itinerant}. In MnPSe$_3$, for instance, the spin-flop field is an order of magnitude smaller than in the sulfide, reflecting reduced anisotropy and a weaker exchange field~\cite{nair2024spin,calder2021magnetic}, in close analogy to the trends we infer for CrPS$_4$. Taken together, this analysis points to a route for ligand engineering by substituting S with Se or Te, or tuning P/S coordination, to directly affect exchange and anisotropy, and thus the spin-flop and spin-flip fields~\cite{basnet2022controlling,tartaglia2020accessing,yun2023ligand}.